\documentclass[12pt]{iopart}%{amsart}
\usepackage{amssymb,latexsym,graphicx,cite}
\newtheorem{theorem}{Theorem}
\newtheorem{lemma}[theorem]{Lemma}

\def\I{\Bbb{I}}
\def\R{\Bbb{R}}
\def\Z{\Bbb{Z}}
\def\G{\Gamma}
\def\qed{$\square$}
\def\H{\mathcal{H}}

\begin{document}
\title{On the location of spectral edges in $\Z$-periodic media}
\author{Pavel Exner$^1$, Peter Kuchment$^2$ and Brian Winn$^3$}
\address{$^1$ Department of Theoretical Physics,
Nuclear Physics Institute, Academy of Sciences, 250 68 \v{R}e\v{z}
near Prague, Czech Republic}
\address{$^2$ Mathematics Department, Texas A\&M University,
College Station, Texas 77843-3368, USA}
\address{$^3$   School of Mathematics,
Loughborough University, Loughborough,
Leicestershire LE11 3TU, UK}
\begin{center}
This paper is dedicated to the memory of P. Duclos.
\end{center}
\begin{abstract}
Periodic $2$nd order ordinary differential operators on $\R$ are
known to have the edges of their spectra to occur only at the
spectra of periodic and antiperiodic boundary value problems. The
multi-dimensional analog of this property is false, as was shown
in a 2007 paper by some of the authors of this article. However,
one sometimes encounters the claims that in the case of a single
periodicity (i.e., with respect to the lattice $\Z$), the $1D$
property still holds, and spectral edges occur at the periodic and
anti-periodic spectra only. In this work we show that even in the
simplest case of quantum graphs this is not true. It is shown that
this is true if the graph consists of a $1D$ chain of finite
graphs connected by single edges, while if the connections are
formed by at least two edges, the spectral edges can already occur
away from the periodic and anti-periodic spectra.
\end{abstract}
\submitto{\JPA}

\maketitle

\section{Introduction}
The following frequently used feature of spectra of periodic second order ordinary differential operators in $L^2(\R)$ is well known and follows from the standard Floquet theory \cite{MagnusWinkl,Eastham}: all edges of the spectral bands of such an operator on the whole axis can be found by considering the spectra of periodic and antiperiodic boundary value problems. In the multi-dimensional case, the role of the periodic and anti-periodic problems is played by symmetry points of the Brillouin zone. Thus, until recently, there had been a rather wide spread belief that these were the only possible locations of the spectral edges. This belief was in most cases confirmed by numerics done for crystalline or photonic crystal media. It was, however, the experience of numerical analysts that in general this claim should be incorrect. This was shown analytically to be indeed the case in \cite{HKSW}. However, after that, claims have been encountered that in the case of a single periodicity (i.e., with respect to the lattice $\Z$), the spectral edges do occur at the periodic and anti-periodic spectra only. In this work we show that even in the simplest case of quantum graphs this is not true. Namely, it is shown in Theorem \ref{T:main} that this is indeed true if the graph consists of a $1D$ chain of finite graphs connected by single edges, while if the connections are formed by two or more edges, the spectral edges can already occur away from the periodic and anti-periodic spectra. Thus, one has to be careful drawing conclusions about $1D$-like spectral behavior when the medium is $1D$-periodic.

\begin{figure}[ht!]
\begin{center}
\setlength{\unitlength}{2cm}
\begin{picture}(5,1)
\put(0.0,0.0){\includegraphics[width=10cm]{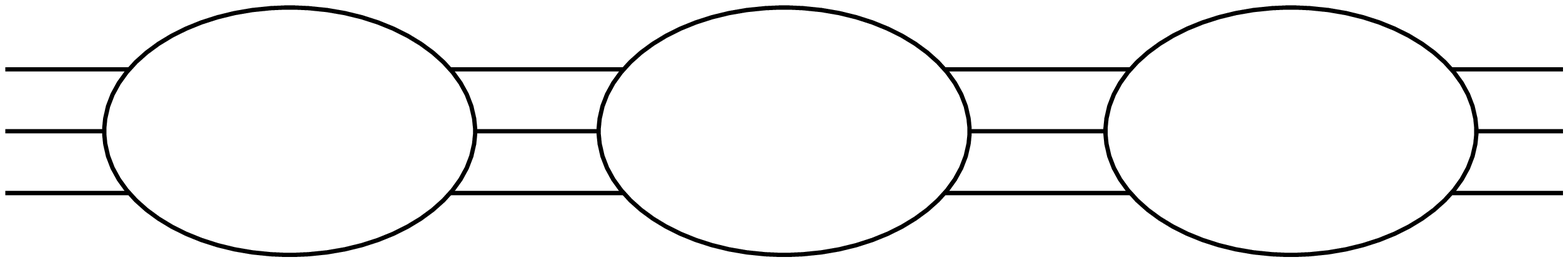}}
\put(0.8,0.35){$\Gamma_{j-1}$}
\put(2.4,0.35){$\Gamma_j$}
\put(4.0,0.35){$\Gamma_{j+1}$}
\end{picture}
\end{center}
\caption{A $\Z$-periodic ``chain'' graph $G$ with three connecting edges.}\label{F:chain}
\end{figure}

We will be considering a graph $G$ that is an infinite ``chain'' of identical copies $\G_j$ of a graph $\G$, consecutive copies being connected by $m$ edges. E.g., Fig. \ref{F:chain} shows such a chain graph with $m=3$ connecting edges. Here the internal structure of the repeated copies $\G_j$ is not important and thus is not revealed.
The graph $G$ is equipped with the natural action by the shifts of the group of integers $\Z$. We will denote this action as $(n,x)\in \Z\times G \mapsto \tau_n x \in G$.

We will assume $G$ to be a quantum graph (see, e.g., the survey \cite{KuAGA} and references therein). In other words, each edge $e$ is assigned a finite length $0<l_e<\infty$ in such a way that it is invariant with respect to the $\Z$-action. Correspondingly, each edge $e$ is provided with a ($\Z$-periodic on the whole $\G$) coordinate function $x_e$ (or just $x$), which identifies $e$ with the segment $[0,l_e]$. Moreover, an operator (Hamiltonian) $\H$ is defined, which acts as the (negative) second derivative $-\rmd^2/\rmd x^2$ on each edge, accompanied with the conditions at the vertices, which require the functions from its domain to be continuous, and their outgoing derivatives along edges to sum to zero at each vertex (the so called Neumann or Kirchhoff vertex conditions \cite{KuAGA}). We define the domain of $\H$ as consisting of all functions $f$ on $G$ such that
\begin{enumerate}
  \item $f\in H^2(e)$ for each edge $e$, where $H^2(e)$ is the standard Sobolev space of order $2$ on the segment $e$ (e.g., \cite{KuAGA}),
  \item
  \begin{equation}\label{E:h2cond}
  \sum\limits_e \|f\|^2_{H^2(e)}<\infty,
  \end{equation}
  \item $f$ satisfies at each vertex $v$ the vertex conditions specified above:
      \begin{equation}\label{E:Kirchhoff}
        f \mbox{ is continuous at }v \mbox{ and }
        \sum\limits_e \frac{\rmd f}{\rmd x_e}(v)=0,
      \end{equation}
      where the sum is taken over all edges adjacent to the vertex $v$ and derivatives are taken in the directions away from $v$.
\end{enumerate}
It is known that this unbounded operator $\H$ in $L^2(G)$ is self-adjoint \cite{KuAGA}.

Due to the $\Z$-periodicity of $\H$, the graph version of the standard Floquet theory applies (see, e.g., \cite{Eastham,MagnusWinkl,RS,Ku93,Ku01} for the Floquet theory for differential equations, and references in \cite{KuAGA} for its graph version). It implies the following features that we will need to use. Let $k\in [-\pi,\pi]$ be a {\bf quasi-momentum} (the name comes from the solid state physics, e.g. \cite{AM}). We consider the Floquet Hamiltonian operator $\H (k)$ that is defined similarly to $\H$, with the exception that the second condition (\ref{E:h2cond}) is replaced by the automorphicity (Floquet, cyclic) condition
\begin{equation}\label{E:h2cond_k}
  f(\tau_nx)=\rme^{\rmi kn}f(x) \mbox{ for any }(n,x)\in \Z\times G.
  \end{equation}

Notice that when $k=0$, condition (\ref{E:h2cond_k}) reduces to the $\Z$-periodicity of $f$, while $k=\pm \pi$ produces the anti-periodic condition. These two cases play an important role in our considerations.

The spectrum of $\H (k)$ is discrete and, numbered in non-decreasing order (counting multiplicity), consists of the eigenvalues $\lambda_j(k), \quad j=1,2,\dots$. The functions $\lambda_j(k)$ are called the {\bf band functions}. The closed interval $I_j$, which is the image of the $j$-th band function, $I_j:=\lambda_j([-\pi,\pi])$, is called the $j$-th {\bf band}. The multiple-valued function
$$
\lambda(k):=\{\lambda_j(k)\}
$$
is said to be the {\bf dispersion relation} and its graph $B$ is called the {\bf Bloch variety}. The level sets of the dispersion relation are often called {\bf Fermi surfaces} (although in physics this term is used for one specific level \cite{AM}).

The following result is well known both in ODE and graph situations:

\begin{theorem}\label{T:Floquet}(see \cite{RS,Eastham,KuAGA,Ku01,Ku93,Titchmarsh} and references therein) \indent
 \begin{enumerate}
 \item The band functions $\lambda_j(k)$ are continuous and piece-wise analytic.
 \item The Bloch variety is an analytic set, i.e. it can be described as the set of all zeros of an analytic function $\phi(k,\lambda)$.
 \item The spectrum of the operator $\H$ in $L^2(G)$ consists of the union of the bands:
\begin{equation}\label{spectr}
    \sigma(\H)=\bigcup\limits_{j}I_j=\bigcup\limits_{k\in[-\pi,\pi]} \sigma(\H(k)).
\end{equation}
\end{enumerate}
\end{theorem}

Our goal is to consider in the $\Z$-periodic quantum graph case the validity of the analog of the following classical theorem:

\begin{theorem}\label{T:ODE}(see \cite{RS,Eastham,Ku01,Ku93,MagnusWinkl,Titchmarsh} and references therein) In the ODE case (i.e., when the graph $G$ is just the straight line), the endpoints of the bands $I_j=\lambda_j([-\pi,\pi])$ (i.e., the extrema of the band functions) are attained at the points $k=0,k=\pm \pi$. In other words, the spectra of the periodic and anti-periodic problems provide the ends of the bands of the spectrum.
\end{theorem}

\section{The main result}
We will now discuss the validity of an analog of Theorem \ref{T:ODE} in the quantum graph case.
\begin{theorem}\label{T:main} Let $G$ be a $\Z$-periodic ``chain'' graph $G$ with $m$ connecting edges (Fig. \ref{F:chain}) and $\H$ be the corresponding Hamiltonian operator in $L^2(G)$. Then
\begin{enumerate}
  \item If $m=1$, the endpoints of the bands $I_j=\lambda_j([-\pi,\pi])$ (i.e., the extrema of the band functions) are attained at the points $k=0,k=\pm \pi$ (although, they might be attained at some other points as well). In other words, the spectra of the periodic and anti-periodic problems provide the ends of the bands of the spectrum.
  \item If $m>1$, this is not always true.
\end{enumerate}
\end{theorem}
%{\bf Proof}

The first statement of the theorem essentially claims that in terms of the spectral band edges, the chain quantum graphs with single connecting edges behave similarly to ODEs.

\begin{figure}[ht!]
\begin{center}
\setlength{\unitlength}{2cm}
\begin{picture}(3,1)
\put(0.0,0.0){\includegraphics[width=6cm]{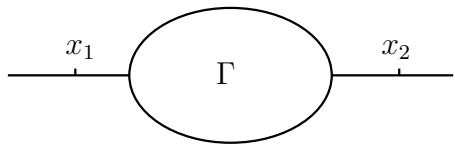}}
\put(0.4,0.6){$x_1$}
\put(2.5,0.6){$x_2$}
\put(1.4,0.4){$\Gamma$}
\end{picture}
\end{center}
\caption{One link of a chain graph with $m=1$, where the two marked points are related by the unit shift: $x_2=\tau_1 x_1$. The internal structure of the link $\G$ is irrelevant and not shown.}\label{F:chain1}
\end{figure}

We start with the following crucial auxiliary statement:
\begin{lemma}\label{L:dimension}\indent
\begin{enumerate}
\item If some value $\lambda$ is attained by the band functions $\lambda_j(k)$ at more than two points $k$ in the segment $(-\pi,\pi]$, then there is a constant branch $\lambda(k)\equiv \lambda$ for all $k$, and thus this value is attained at all points of the segment.

\item The set $D$ of all such values $\lambda$ is discrete (possibly empty).
\item If $\lambda\notin D$, then in a neighborhood of this value all band functions are strictly monotonic on $[0,\pi]$.
\end{enumerate}
\end{lemma}
{\bf Proof of the Lemma.}\phantom{M}
Figure \ref{F:chain1} shows a link of a chain graph with $m=1$.
%
%\begin{figure}[ht!]
%\begin{center}
%\scalebox{.5}{\includegraphics{chain1.eps}}
%\end{center}
%\end{figure}
Suppose that a value $\lambda$ is taken by band functions $\lambda_j(k)$ at more than two points $k\in [-\pi,\pi]$. This means that the space of (bounded) solutions of the equation $(\H-\lambda \I)u=0$ on $G$ is more than two-dimensional. Hence, there exists a non-trivial solution $u$ that vanishes with its first derivative at a point $x_1$ on the connecting edge (see Fig. \ref{F:chain1}). Since the equation is of the second order on each edge, we conclude that $u$ vanishes on the whole edge containing $x_1$. Suppose that $u$ is not identically equal to zero on the part of the graph to the right of $x_1$ (the case when this happens to the left can be treated in a similar manner). Re-defining $u$ as being equal to zero everywhere to the left of $x_1$, one has a legitimate bounded solution, which we will denote by $u$ again. Then the linear hull of the right shifts of $u$ is infinitely dimensional. However, the Floquet theory (e.g., \cite{Ku_graphs2}) implies that then the ``Fermi surface'' of all points  $k\in [0,\pi]$ where $\lambda_j(k)=\lambda$ for some $j$ must also be infinite. Then the analytic structure (see Theorem \ref{T:Floquet}) of the Bloch variety
$$
B=\{(k,\lambda)\mid \lambda_j(k)=\lambda \mbox{ for some }j\}
$$
implies that it contains the flat branch $\lambda_j(k)\equiv\lambda$ and that this can occur only at a discrete set $D$ of values of $\lambda$. This proves the first two statements of the Lemma.

Consider now a value $\lambda\notin D$. Due to the invariance of the operator $\H$ with respect to the complex conjugation (which corresponds to the time reversal symmetry of the system), the band functions $\lambda_j(k)$ are even: $\lambda_j(-k)=\lambda_j(k)$. This, together with the first statement of the Lemma and the assumption that $\lambda\notin D$ implies that all values near $\lambda$ are attained by the (continuous) function $\lambda_j(k)$ only once on $[0,\pi]$. Thus, this function is monotonic there, which proves the last claim of the Lemma.
\qed

\noindent
{\bf Proof of Theorem \ref{T:main}.}\phantom{M} Suppose that $\lambda$ is a band edge, i.e. an extremum of a band function $\lambda_j(k)$. If $\lambda\in D$, where $D$ is the set defined in the Lemma, then the statement (i) of the Lemma claims that this value $\lambda$ is attained at all values of quasi-momentum $k$, in particular for $k=\pi$ and $k=0$ that correspond to the anti-periodic and periodic problems.

Let now $\lambda\notin D$. Then the statement (iii) of the Lemma implies that the corresponding value of $k$ cannot be in the interior $(0,\pi)$ of the segment $[0,\pi]$. Thus, either $k=0$ and $\lambda$ belongs to the spectrum of the periodic problem, or  $k=\pi$ and $\lambda$ belongs to the spectrum of the anti-periodic problem. This proves the first statement of the theorem.

Let us now construct an example with $m=2$, where some of the band edges are attained not on the spectra of the periodic and anti-periodic problems. We start with the $\Z^2$-periodic graph considered in \cite[Fig. 1]{HKSW}. Its fundamental domain $W$ (with the dotted line boundary) is shown in Figure \ref{fig:quantumfd} below.
\begin{figure}[ht!]
\begin{center}
\setlength{\unitlength}{4cm}
\begin{picture}(1,1)
\put(0.0,0.0){\includegraphics[width=4cm]{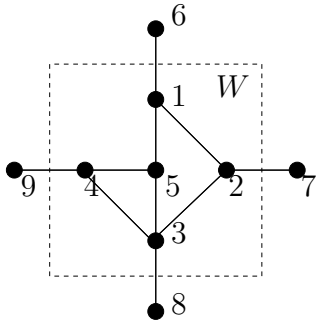}}
\put(0.26,0.41){$4$} \put(0.53,0.41){$5$} \put(0.74,0.41){$2$}
\put(0.55,0.257){$3$} \put(0.55,0.71){$1$} \put(0.7,0.74){$W$}
\put(0.05,0.41){$9$} \put(0.55,0.02){$8$} \put(0.98,0.41){$7$}
\put(0.55,0.98){$6$}
\end{picture}
\caption{The graph $\Gamma$ with fundamental region $W$.}
\label{fig:quantumfd}
\end{center}
\end{figure}
One imagines the whole graph as obtained by tiling the plane with the $\Z^2$-shifted copies of $W$. The lengths of all edges are assumed to be equal to $1$. The two basic period vectors are $v_1$ connecting vertices $4$ to $7$ and $v_2$ connecting $3$ to $6$.

It was shown in \cite{HKSW} that some of the band functions of the discrete Laplace operator on this graph attain their extremal values strictly inside the reduced Brillouin zone $[0,\pi]^2$. It was then shown that this implies the same feature for the quantum graph operator $\H$.

\begin{figure}[ht!]
\begin{center}
\setlength{\unitlength}{5cm}
\begin{picture}(1,1)
\put(0.0,0.0){\includegraphics[width=5cm]{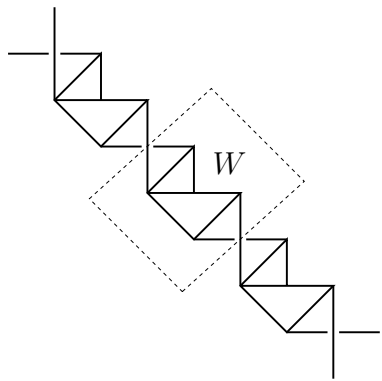}}
\put(0.55,0.55){$W$}
%\scalebox{.4}{\includegraphics{Zgraph.eps}}
\end{picture}
\end{center}
\caption{The $\Z$-periodic graph obtained by folding the graph of Fig. \ref{fig:quantumfd}.}\label{F:Zgraph}
\end{figure}

We obtain our example by folding this graph along the period vector $v:=v_1+v_2$. In other words, the points $x$ and $y$ such that $y=v+x$ are identified. This leads to the $\Z$-periodic graph $G$ shown in Fig. \ref{F:Zgraph}.
This graph is clearly $\Z$-periodic and has $m=2$ (the two skew edges connecting the adjacent copies of the fundamental domain $W$).

\begin{figure}[ht!]
\begin{center}
\scalebox{.6}{\includegraphics{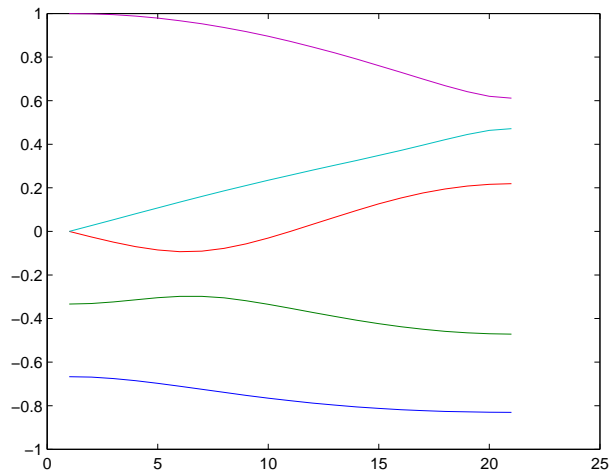}}
\end{center}
\caption{The graphs over $[0,\pi]$ of the five band functions of $\Delta$.}\label{F:plot}
\end{figure}

Let us consider the version of the discrete Laplace-Beltrami operator $\Delta$ on functions defined on the vertices of $G$, which adds the values at all neighboring vertices,
weighted by the reciprocals of the square roots of the degrees
(valences):
$$
(\Delta f)(v):=\frac{1}{\sqrt{d_v}}\sum\limits_{u\sim v}\frac{1}{\sqrt{d_u}}f(u),
$$
where $u$ and $v$ are vertices, $d_u$ is the degree of $v$, and the sum is taken over all vertices $u$ adjacent to $v$ (in the paper \cite{HKSW} this operator was denoted $L$).
For instance, for a function $f$ defined on the vertices of the graph in Fig. \ref{fig:quantumfd} one has
$$
(\Delta f)(5)=\frac1{\sqrt{3}}\left(\frac1{\sqrt{3}}f(1)+\frac12f(3)+
\frac1{\sqrt{3}}f(4)\right),$$
and
$$ (\Delta f)(3)=\frac12\left(\frac1{\sqrt{3}}f(2)+\frac1{\sqrt{3}}f(4)+
\frac1{\sqrt{3}}f(5)+\frac1{\sqrt{3}}f(8)\right), \mbox{{\it etc}.}
$$
Then direct calculation shows that for any quasi-momentum $k$, the Floquet operator $\Delta(k)$ is given by the following $5\times 5$ matrix:
\begin{equation}\label{E:matrix}
  \Delta(k)=  \frac1{\sqrt{3}}\left(
      \begin{array}{ccccc}
        0 & 3^{-1/2} & \rme^{\rmi k}/2 & 0 & 3^{-1/2} \\
        3^{-1/2} & 0 & 1/2 & \rme^{\rmi k}/3^{1/2} & 0 \\
        \rme^{-\rmi k}/2 & 1/2 & 0 & 1/2 & 1/2 \\
        0 & \rme^{-\rmi k}/3^{1/2} & 1/2 & 0 & 3^{-1/2} \\
        3^{-1/2} & 0 & 1/2 & 3^{-1/2} & 0 \\
      \end{array}
    \right)
\end{equation}
The plot of the five eigenvalues of $\Delta(k)$ for $k=(l-1)\pi/20,l=1\ldots21$ is shown in Fig. \ref{F:plot}.

One sees that the minimum of the third branch, as well as the maximum of the second one are attained inside the segment, and thus not at the points $k=0$ or $k=\pi$. This numerical observation can be confirmed, as in \cite{HKSW}, by a somewhat tedious explicit calculation of the dispersion relation and elementary analysis of its behavior. Again, as in \cite{HKSW}, one shows that this fact about the band functions of the discrete operator $\Delta$, in the case of all edges having equal length implies the same conclusion for the quantum graph Hamiltonian $\H$ on $G$.
This provides the required counter-example for $m=2$. It is easy to figure out that now one can create examples of this kind for any $m>1$, just by adding an unrelated structure to the example above.

Thus, the Theorem is proven.
\qed

\section{Remarks}

\begin{enumerate}
  \item The set $D$ of the exceptional values in Theorem \ref{T:main} constitutes the pure point part of the spectrum of the operator $\H$. In the case of periodic ODEs and 2nd order PDEs it is known that this part of the spectrum is empty (see, e.g., the surveys \cite{Ku01,BirSu}) and the whole spectrum is absolutely continuous (the singular continuous part is always absent in the periodic situation \cite{Ku93}).

      The reason for appearance of the pure point spectrum in the graph case is that it corresponds to compactly supported eigenfunctions \cite{Ku_graphs2}. Such compactly supported solutions are known not to exist for the second order elliptic operators.
  \item Various spectral investigations and applications of periodic quantum graphs can be found, for instance, in \cite{AEL,Exner_KrPen,Exner_rectang,volume,kuchment:91,Ku_graphs2,KuchPost,KuchVain3,Oleinik} (see also references in these works).
  \item The first statement of Theorem \ref{T:main} should be understood correctly: when $m=1$, the extrema are attained at $k=0$ and $k=\pi$. However, as the proof shows, in contrast with the ODE case, it can be also attained elsewhere, if a flat branch of dispersion relation is present.
\end{enumerate}

\section*{Acknowledgments}

The authors express their gratitude to G. Berkolaiko for his suggestion to discuss the $\Z$-periodic case and for useful comments about the proofs. The work of the first author was supported in part by the Czech
Ministry of Education, Youth and Sports within the project LC06002. The work of the second author was partially supported by the KAUST grant KUS-CI-016-04 through the Inst. Appl. Math. Comp. Sci. (IAMCS) at Texas A\&M University.
The third author has been financially supported by the National Sciences
Foundation under research grant DMS-0604859.
The authors are grateful to these agencies for the support.

\end{document}